\documentclass[12pt]{iopart}

\usepackage{graphicx} 
\usepackage[backend=biber,style=authoryear,sorting=nyt]{biblatex}
\usepackage{hyperref}
\usepackage{comment}

\usepackage{subcaption}  

\usepackage{array}
\usepackage{booktabs}
\usepackage[a4paper, top=2.5cm, bottom=2.5cm, left=2.5cm, right=2.5cm]{geometry}
\usepackage{iopams} 

\usepackage{xcolor}
\usepackage{orcidlink}

\begin{document}
\pagestyle{plain}


\title{Machine Learning-Based Modeling of the Anode Heel Effect in X-ray Beam Monte Carlo Simulations}

\author{Hussein Harb\textsuperscript{1}\textsuperscript{*}\orcidlink{0009-0004-9615-6831}, 
Didier Benoit\textsuperscript{1}, 
Axel Rannou\textsuperscript{1}, 
Chi-Hieu Pham\textsuperscript{1}, 
Valentin Tissot\textsuperscript{2},
Bahaa Nasr\textsuperscript{1,2}, and Julien Bert\textsuperscript{1,2}\orcidlink{0000-0002-2756-942X}}

\address{\textsuperscript{1} LaTIM, University of Brest, INSERM UMR1101, Brest, France}
\address{\textsuperscript{2} Brest University Hospital, France}
\address{\textsuperscript{*} Author to whom any correspondence should be addressed.}

\ead{hussein.harb@univ-brest.fr}


\begin{abstract}
\textit{Objective.}  
To develop a machine learning-based framework for accurately modeling the anode heel effect in Monte Carlo simulations of X-ray imaging systems, enabling realistic beam intensity profiles with minimal experimental calibration.
\textit{Approach.}  
Multiple regression models were trained to predict spatial intensity variations along the anode–cathode axis using experimentally acquired weights derived from beam measurements across different tube potentials. These weights captured the asymmetry introduced by the anode heel effect. A systematic fine-tuning protocol was established to minimize the number of required measurements while preserving model accuracy. The models were implemented in the OpenGATE 10 and GGEMS Monte Carlo toolkits to evaluate their integration feasibility and predictive performance.
\textit{Main results.}  
Among the tested models, gradient boosting regression (GBR) delivered the highest accuracy, with prediction errors remaining below 5\% across all energy levels. The optimized fine-tuning strategy required only six detector positions per energy level, reducing measurement effort by 65\%. The maximum error introduced through this fine-tuning process remained below 2\%. Dose actor comparisons within Monte Carlo simulations demonstrated that the GBR-based model closely replicated clinical beam profiles and significantly outperformed conventional symmetric beam models.
\textit{Significance.}  
This study presents a robust and generalizable method for incorporating the anode heel effect into Monte Carlo simulations using machine learning. By enabling accurate, energy-dependent beam modeling with limited calibration data, the approach enhances simulation realism for applications in clinical dosimetry, image quality assessment, and radiation protection.
\end{abstract}


\section{Introduction}
X-ray imaging plays a pivotal role in both diagnostic and interventional medicine. Due to its ionizing nature, optimizing imaging systems, protocols, and reconstruction methods is crucial to maintaining high image quality while minimizing patient exposure. In this context, Monte Carlo (MC) simulation remains an indispensable tool in medical physics, offering unmatched accuracy in modeling photon-matter interactions and dose deposition \parencite{papadimitroulas2017dosimetry, sarrut2022}. This approach allows the simulation of the entire X-ray tube system \parencite{seibert2004x}, enabling the generation of highly realistic photon spectra and spatial distributions. Nevertheless, MC simulations are computationally intensive. 

A common approach to reduce computational costs is to divide the simulation into two phases: (1) a simulation of the X-ray tube alone, generating a phase space file, and (2) a patient-dependent phase where this file is reused, eliminating the need for repeated tube simulations. However, for simulations that require a large number of particles and extensive statistics, the phase space file can reach hundreds of gigabytes, making the simulation impractical. Generally, a virtual source is preferred. This refers to an analytical mathematical model that represents the distribution and correlation of particles in the phase space \parencite{sarrut2019generative}. Such model requires only few parameters saving stored memory and can rapidly produce particle distributions on demand. The main drawback is that such models are often an approximation and only represent the predominant characteristics of the source.

For example, the anode heel effect of the X-ray tube is rarely considered. It affects the beam intensity asymmetry arising from the angled geometry of the X-ray tube anode. This effect causes X-rays directed toward the cathode side to exhibit higher intensities, while those toward the anode side are attenuated due to self-absorption within the anode~\parencite{shin2016anode, behling2018diagnostic}. This asymmetry impacts both dose distribution and image quality, particularly in wide-field imaging applications~\parencite{mould1995early}. Neglecting the anode heel effect in MC simulations can lead to significant discrepancies between simulated and real measurements~\parencite{prabhu2020production}.

Recent studies have explored various approaches to model the anode heel effect, including empirical adjustments and analytical formulations~\parencite{poludniowski2021spekpy}. Traditional analytical and empirical descriptions of the heel effect~\parencite{seibert2004x, shin2016anode, behling2020x} mainly focus on modeling the energy spectrum of the emitted X-rays by accounting for attenuation within the anode material. These models typically integrate the heel effect as an energy-dependent correction during spectrum generation, without explicitly describing the spatial fluence variation along the anode–cathode axis. Advances in artificial intelligence (AI) present a promising solution for capturing complex, non-linear phenomena like the anode heel effect. Machine learning models, trained on experimental data, can generalize across a wide range of energy levels and beam geometries, offering a scalable alternative to traditional methods \parencite{keal2021dosecalc}.

For this, we propose a machine learning-based framework to model the anode heel effect with high fidelity and minimal calibration effort. A key component of the proposed model is a systematic fine-tuning strategy that enables rapid adaptation to different X-ray systems using only a limited number of measurements. Specifically, we show that acquiring dose data at just six measurements per energy level  is sufficient to accurately reconstruct the beam profile, significantly reducing experimental efforts. The model is trained using experimental data combined with a Gradient Boosting Regression algorithm to predict spatial beam intensity distributions across energy levels. The proposed anode heel effect model was subsequently integrated into the OpenGATE 10 \parencite{sarrut2025new, krah2025new} and GGEMS \parencite{Bert2013} MC toolkits, enabling precise and realistic simulation of clinical X-ray beams. By addressing both accuracy and adaptability, this approach provides a robust and scalable solution for improving dosimetry, image quality, and radiation safety in diverse medical imaging environments.

\section{Materials and Methods}
\label{section2}

In this study, we modeled a virtual source that replicate the characteristics and behavior of a real X-ray tube. This tube consists of a cathode that emits electrons through thermionic emission and a tungsten anode where these electrons are accelerated under high voltage and converted into X-ray photons via bremsstrahlung and characteristic interactions \parencite{behling2020x, zink1997x}. The anode's angled geometry plays a critical role in producing the non-uniform intensity distribution of the beam, contributing to the anode heel effect, as shown in Figure\ref{anode}.

\subsection{Standard X-ray tube virtual source model}
A virtual source model consists of mathematically modeling the emission of photons in order to simplify and speed the  simulation. In MC simulations, for each randomly emitted particle, the parameters of the photon have to be calculated following the considering source. Each photon $\gamma(E, \mathbf{p}, \mathbf{d})$, is defined by its energy $E$, its position $\mathbf{p} = \{p_x, p_y, p_z\}^T$, and its momentum (direction) $\mathbf{d} = \{d_x, d_y, d_z\}^T$. In the context of X-ray tube simulation, the position $\mathbf{p}$ is defined by the emission area on the anode, which corresponds to a rectangular surface measured in millimeters (mm). For each new particle, this position is determined randomly and uniformly over this surface, $\mathbf{p} \sim \mathcal{U}(S)$, where $\mathcal{U}$ represents the uniform distribution, and $S$ denotes the surface.

\begin{figure}[]
    \centering
    \begin{subfigure}[t]{0.49\linewidth} 
        \centering
        \includegraphics[width=\linewidth]{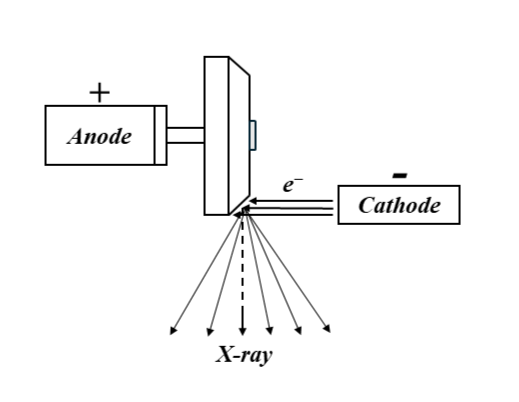}
        \caption{}
        \label{anode}
    \end{subfigure}
    \hspace{0.01\linewidth} 
    \begin{subfigure}[t]{0.40\linewidth} 
        \centering
        \includegraphics[width=\linewidth]{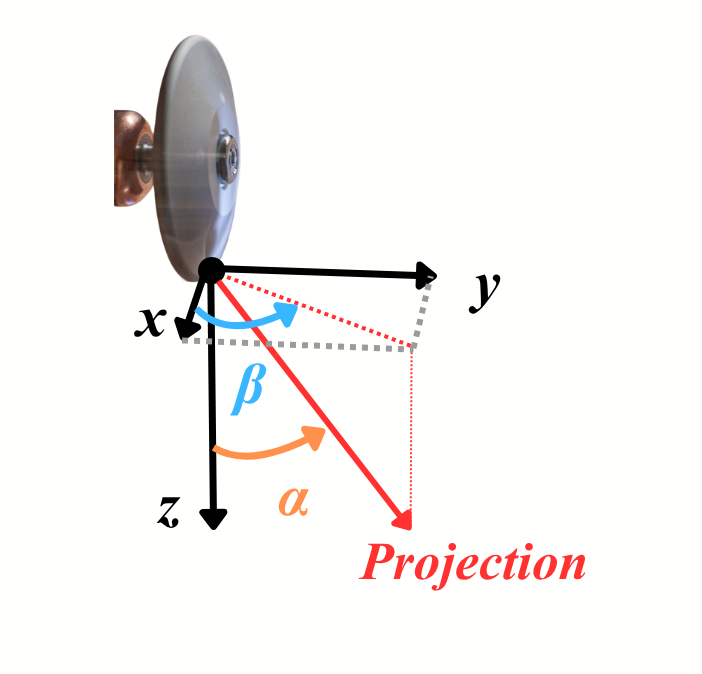}
        \caption{}
        \label{anode2}
    \end{subfigure}
    \caption{(a) Simplified schematic of the X-ray tube showing the anode–cathode orientation and the direction along which dose measurements were performed. (b) Schematic representation of the virtual X-ray source model, illustrating the emission angles $(\alpha: altitude, \beta: azimuthal)$ defining photon projection in the simulation. }
    \label{anodeee}
\end{figure}

The particle's momentum is defined by the Euler angles $(\alpha: altitude, \beta: azimuthal)$ of the particle emission. The geometry of the virtual source and the definition of Euler angles are illustrated in Figure \ref{anode2}. In the standard model, particles are emitted following an angular uniform distribution. In this case the random sampling should not be done over the altitude angle $\alpha$, but rather over its cosine $\cos(\alpha)$. This can be expressed as:

\begin{equation}
	\frac{d\Omega}{d\cos \alpha} = 1
\end{equation}

where $d\Omega$ is the differential solid angle and $\cos(\alpha)$ is the cosine of the polar angle $\alpha$. Therefore, the momentum direction $\mathbf{d}$ is sampled randomly with a probability distribution proportional to $\mathcal{U}(\cos \alpha)$. The azimuthal angle $\beta$ is also sampled randomly and uniformly between $0$ and $2\pi$: $\beta \sim \mathcal{U}(0, 2\pi)$. However, the angle of the beam's aperture $\mathcal{A}$ must also be taken into account. The angle $\alpha$ is then obtained as follows:

\begin{equation}
	\alpha = \cos^{-1} \left( 1 + \xi \cdot \left[\cos(\mathcal{A}) - 1 \right] \right)
\end{equation}

where $\xi$ is a uniform random number between $[0, 1)$. Based on the two angles $ \alpha $ and $ \beta $, the direction of the particle $ \mathbf{d} $ can be defined as follows:

\begin{equation}
	\mathbf{d} = 
	\left[ 
	\begin{array}{c}
		\cos \beta \cdot \sin \alpha \\
		\sin \beta \cdot \sin \alpha \\
		\cos \alpha
	\end{array}
	\right]
\end{equation}

Most of the time, the tube is also oriented around the patient, then a transformation including rotation and translation must be applied to the particle:

\begin{equation}
		\left\{ 
		\begin{array}{l}
			\mathbf{p'} = \mathbf{T}_{xyz}\mathbf{p} \\
			\mathbf{d'} = \mathbf{R}_{\psi\theta\phi}\mathbf{d}
		\end{array}
		\right.
\end{equation}

where $ \mathbf{p'} $ is the final particle emitting position, $ \mathbf{d'} $ is the final particle momentum, $\mathbf{R}$ and $\mathbf{T}$ are the rotation and translation matrices of the X-ray tube respectively. Collimation can also be applied, whether to achieve a fan beam in diagnostic CT or to reduce the field of view in fluoroscopy. 

Two strategies are commonly used. The first involves defining the collimator as a patient-dependent component, as it can be tailored to the specific clinical case. In this strategy, the collimator is simulated each time using MC methods, which also allows for the calculation of beam scattering (penumbra). This approach is particularly useful when filters, such as wedges or bowtie filters, are present within the beam. If the collimation is fixed and no filters are used, a second strategy involves employing a ray-tracing method within the virtual source model. Essentially, the particle path is checked to determine if it hits the collimator. If it does, a new particle is emitted. This process is repeated until a particle successfully exits the source, similar to a rejection sampling method. Alike source and collimation modeling strategies have been previously described in Monte Carlo and generative modeling frameworks \parencite{fix2004monte, sarrut2019generative}.

Regarding the determination of the particle’s energy, a realistic emission spectrum is used. Existing model, such as the TASMIP model \parencite{boone_accurate_1997} or, more recently, CASIM \parencite{omar_model_2018}, which is the default model in the Python library SpekPy \parencite{poludniowski2021spekpy}, can be used to provide spectrum histogram $f(E_i)$ based on the tube parameters, such as voltage, anode material, or filtration layers. The particle energy is then randomly sampled by using the cumulative distribution function (CDF) obtained by normalizing the histogram $(\sum f(E_i)=1)$, and by cumulating the bin value as follow:

\begin{equation}
	F(E_k) = \sum_{i=1}^{k} f_{\mathrm{norm}}(E_i)
\end{equation}

where $F(E_k)$ represents the cumulative probability up to the bin $k$. Given a random number $\xi \sim \mathcal{U}(0, 1)$, the energy $E_\xi$ is determined by finding the smallest bin index $k$ where:

\begin{equation}
	F(E_k) \leq \xi < F(E_{k+1})
\end{equation}

Since the real spectrum is continuous rather than discrete, the final particle energy is determined using linear interpolation.

The present model is developed and validated for reflective tungsten anodes, which are standard in diagnostic X-ray imaging systems. The emission characteristics of transmission targets, such as those used in high-energy radiotherapy accelerators, differ significantly and are beyond the scope of this study.

\subsection{Incorporating the anode heel effect}

The anode heel effect introduces significant variations in beam intensity depending on the emission angle of the particles. To accurately model these variations, angular weighting factors $\mathcal{W}(\alpha, \beta, E)$ are introduced, which depend on both the emission angles and the photon energy. Each value $\mathcal{W}_i$ can be derived either from simulations or experimental measurements by comparing the probability density of emission inferred from image intensities or measured doses. Rather than storing a full multidimensional histogram for every possible configuration, the weights are organized per energy bin. For each energy level, the corresponding angular weight map is transformed into a CDF over the angular domain. During particle generation, the photon energy \(E\) is first drawn according to the X-ray spectrum. Then, the corresponding CDF is used to sample an emission direction \((\alpha, \beta)\), with linear interpolation applied to smooth the transitions between discrete angular bins. Finally, as in the standard Monte Carlo pipeline, the sampled emission angles are converted into particle momentum vectors, accounting for the X-ray tube’s position and orientation in the simulation geometry.

\subsection{Machine Learning-based weight modeling}

To implement the proposed approach, the machine learning model must be trained to predict spatial beam weights as a function of acquisition parameters. This section outlines the input parametrization used for training and describes how the model learns to reproduce reference dose characteristics derived from either measurements or from full MC simulations. The goal is to capture the intensity profile of the beam along the anode-cathode axis, illustrated in Figure \ref{anode}, with high fidelity while minimizing the number of required measurements.

\subsubsection{Model parametrization}

The weights represent the ratio of the measured dose value at a given position to the central beam intensity as explained in the following equation:

\begin{equation} 
W(x) = \frac{D(x)}{D_0} 
\end{equation}

where  \(D_0\) represents the dose value measured at the central axis of the beam, and \(D(x)\) denotes the dose value at a lateral distance from the center. 

The training dataset consists of these weights paired with their corresponding input parameters, including the tube voltage (kVp) and the lateral distance from the beam center (in cm). These parameters serve as the input features for the model.

\subsubsection{Learning the weights with Machine Learning}

To achieve a realistic simulation and cover all parameters of the X-ray tube, it is necessary to determine a significant number of weights, which can be complicated and tedious. When the relationship between parameters is not overly complex and the values are continuous, a mathematical model is often used. The strategy involves fitting a model to the data to determine the coefficients that describe it. This reduces the amount of data needed to store and describe the virtual source. 

However, most of the time, the correlations between parameters are significant, making it difficult to derive a model, especially with complex sources such as those in radiotherapy\parencite{fix2004monte}. The ability of AI to learn a complex model with few coefficients has been explored, particularly for source models in external radiotherapy calculated by MC simulation\parencite{sarrut2019generative}. In this study, we propose using regression models to interpolate the complex behavior of beam weights $\mathcal{W}$ as a function of multiple parameters. Our approach aims to reduce the number of experimental measurements or simulations required to accurately construct the model, improving efficiency without sacrificing fidelity.

\subsection{Cross-machine fine-tuning strategies}
\label{Cross-machine fine-tuning strategies}
A key objective in applying machine learning to model the anode heel effect is to develop a system that can be efficiently adapted to different X-ray machines using a minimal number of experimental measurements. To systematically identify the most effective fine-tuning protocols for cross-machine adaptation, a variety of strategies were designed and evaluated.

To fine-tune the model for a new machine, we used a residual learning approach. In this method, a second model was trained to learn the difference between the original model’s predictions (based on data from Machine 1) and the actual measurements from Machine 2. This approach, known as transfer learning \parencite{5288526, shojaie2022transfer}, allowed us to reuse what the model had already learned from the first machine while adjusting it to the specific characteristics of the second one using correction factors.

Each strategy was assessed using a consistent train–test methodology: the model was fine-tuned on a selected subset of measurement points and then evaluated on the remaining data. This systematic evaluation enabled the identification of fine-tuning approaches that optimally balance measurement effort with predictive accuracy, offering practical guidance for the efficient commissioning of machine-specific beam models.

\subsection{Evaluation}

\subsubsection{Experimental data collection}
\label{Experimental measurements}

\begin{figure}[]
    \centering
    \begin{subfigure}[t]{0.45\linewidth} 
        \centering
        \includegraphics[width=\linewidth]{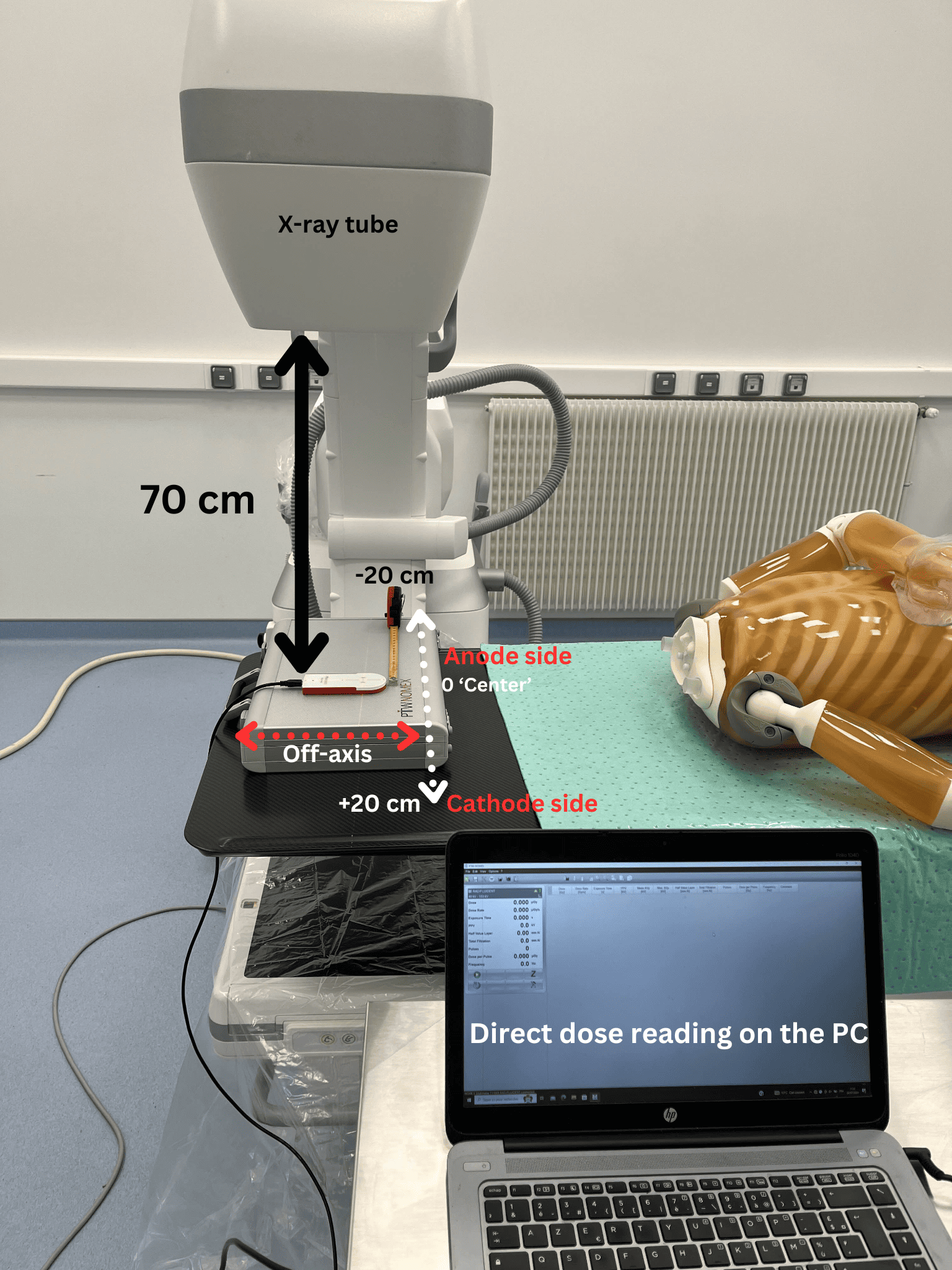}
        \caption{}
        \label{figure_1}
    \end{subfigure}
    \hfill
    \begin{subfigure}[t]{0.45\linewidth} 
        \centering
        \includegraphics[width=\linewidth]{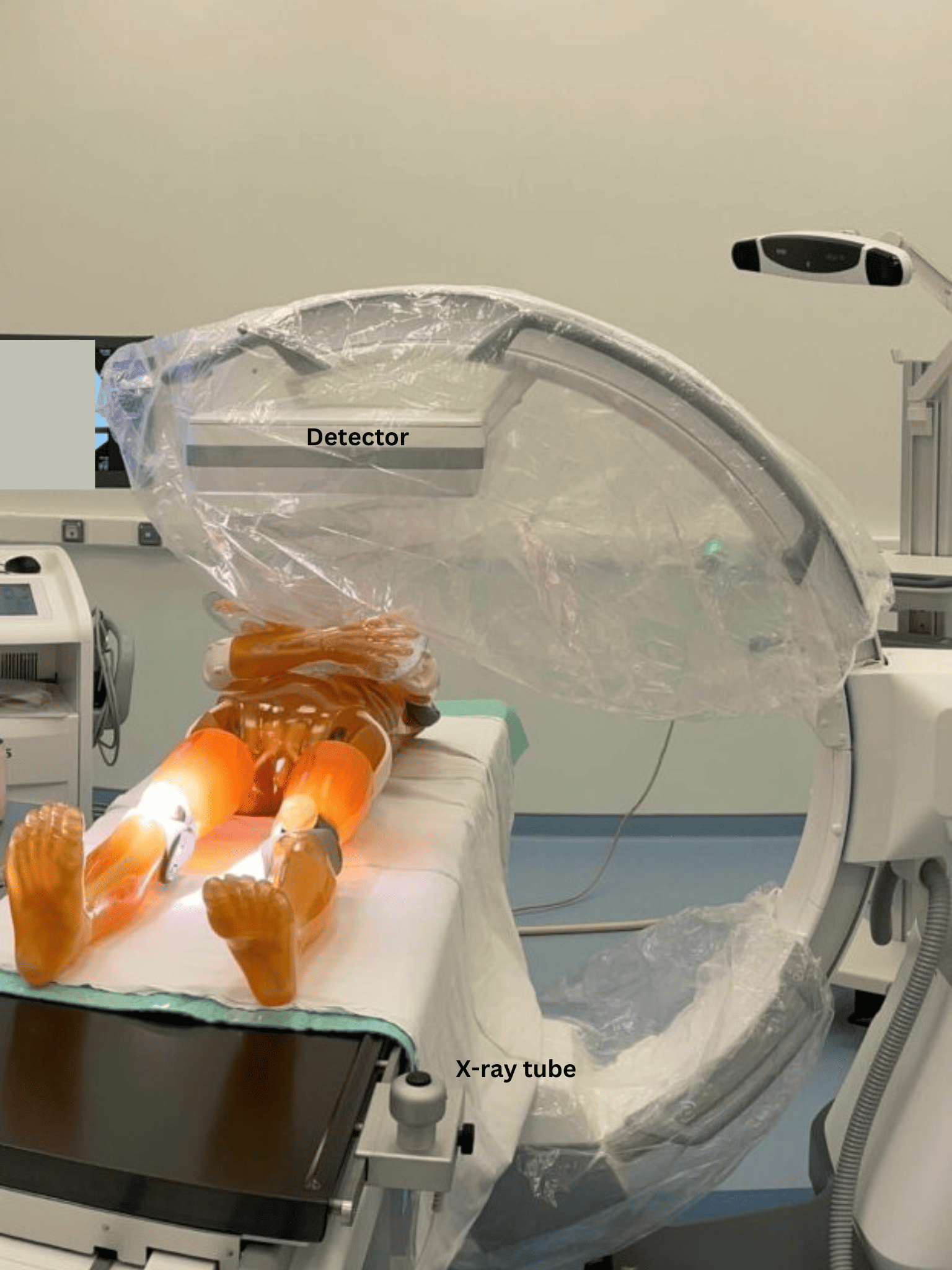}
        \caption{}
        \label{figure1_1}
    \end{subfigure}
    \caption{Overview of the experimental configuration. (a) Experimental setup for measuring X-ray dose values along the anode–cathode axis using a PTW NOMEX Multimeter dosimeter. (b) Posterior–anterior chest X-ray positioning of the anthropomorphic phantom.}
    \label{figure_1_1}
\end{figure}

To better model the system used in our clinical application and demonstrate the interpolation capabilities of machine learning, we determined the weights $\mathcal{W}$ based on experimental measurements in context of interventional radiology application. While MC simulations are a valid alternative, we relied on real measurements using an X-ray machine and a dosimeter to provide ground truth dose values for training the model. The X-ray imaging system utilized in our study was provided by PLaTIMeD\parencite{platimed2024}. The Siemens Cios Alpha system (Siemens Healthcare, Erlangen, Germany)\parencite{siemens2020ciosalpha} was employed, featuring a tungsten (W) anode with a 10° tilt angle, consistent with standard diagnostic X-ray tube designs. This anode geometry induces the characteristic asymmetry in the emitted X-ray intensity profile observed experimentally. Tungsten, owing to its high atomic number and melting point, remains the material of choice for anode targets in diagnostic imaging applications. Dose value measurements were conducted along the anode-cathode axis using a PTW-NOMEX Multimeter dosimeter with measuring error less than 5\% \parencite{ptw_nomex_multimeter}. To ensure measurement consistency, the dosimeter was placed at predefined positions along the axis, with a fixed source-to-detector distance at 70 cm, as shown in figure \ref{figure_1}.

Measurements were performed for tube potentials ranging from 50 to 120 kVp with a 10 kVp step. To account for variations in exposure conditions, dose values were normalized relative to the intensity of the central beam. We conducted a total of 25 measurements per energy level, beginning with the central reference measurement at the intersection of the laser cross "Beam Center". From there, we incrementally shifted by 1 cm in both directions, recording dose values until we reached 9 cm, which marked the edge of the X-ray beam coverage on both sides. To confirm this boundary, we took an additional measurement at 10 cm. Further measurements were performed at 15 cm and 20 cm on both sides to ensure the absence of beam leakage, which could impact the accuracy of the results. Prior to data acquisition, the X-ray system was verified and calibrated through standard quality assurance procedures performed by the hospital’s medical physics department at PLaTIMeD\parencite{platimed2024}, in addition to the maintenance performed by the Siemens Healthineers company, ensuring that the tube output and dosimetric readings were within clinical tolerance limits.

\subsubsection{Machine Learning model evaluation}

A total of six machine learning algorithms were evaluated, including Linear Regression\parencite{yao2014new}, Support Vector Regression (SVR)\parencite{awad2015support}, Multilayer Perceptron (MLP) \parencite{taud2017multilayer}, Random Forest Regressor\parencite{rigatti2017random}, Decision Tree Regressor\parencite{quinlan1986induction}, and Gradient Boosting Regressor (GBR)\parencite{natekin2013gradient}. These models were selected based on their suitability for regression tasks and their ability to model potentially non-linear relationships. The dataset was split into training and validation (80\%) and testing (20\%) subsets to ensure robust model evaluation. Hyperparameter optimization was performed using 5-fold cross-validation to minimize overfitting and improve predictive accuracy. During optimization, the learning rate was varied between 0.001 and 0.1, while the number of estimators ranged from 100 to 1000. The features 'DoseRate', 'Energy', and 'Theta' served as the input variables (X), while 'Weight' was used as the target variable (Y) for training and testing.

Model performance was quantitatively evaluated using the Mean Squared Error (MSE) and the coefficient of determination \( R^2 \), two standard metrics for assessing the accuracy and explanatory power of regression models.

\begin{equation}
	\mathrm{MSE} = \frac{1}{n} \sum_{i=1}^{n} \left( y_i - \hat{y}_i \right)^2,
\end{equation}

where \( y_i \) represents the actual beam weight at a given position and energy, and \( \hat{y}_i \) is the model-predicted value.

\begin{equation}
	R^2 = 1 - \frac{\sum_{i=1}^{n} (y_i - \hat{y}_i)^2}{\sum_{i=1}^{n} (y_i - \bar{y})^2},
\end{equation}

where \( \bar{y} \) is the mean of the observed values. A high \( R^2 \) score reflects the model’s efficacy in accurately predicting beam weights across varying tube potentials and spatial positions.

\subsubsection{Experimental data for cross-machine validation}
\label{Methods:strategies}
 Additional measurements were carried out using a Philips C-arm System 7000–Zenition 70~\parencite{philipsZenition70} at Cavale Hospital, Brest, France, under the same acquisition parameters previously described in Section~\ref{Experimental measurements}. These data were used to investigate multiple strategies for fine-tuning a base model initially trained on measurements from the Siemens system. These strategies were organized into three principal categories, each containing different variants. A variant refers to a distinct configuration for selecting training inputs such as measurement positions or energy levels used to assess how different sampling schemes affect the model’s generalization and prediction accuracy.

\emph{Points-per-energy strategies:} Four variants examined the effect of measurement density by selecting 2, 4, 6, or 8 specific dosimeter positions per energy level ranging from 50 to 120~kVp. The selected points ranged from minimal sampling at the field boundaries ($-8$~cm and $+8$~cm) to more comprehensive spatial distributions.

\emph{Distance pattern strategies:} Seven variants assessed the impact of different spatial sampling patterns. These included boundary-only measurements (\emph{extremes}), quartile-based sampling (\emph{subdivision}), center-only (\emph{assuming symmetry}), edges-only (\emph{capturing heel effect variability}), and systematic even/odd spatial intervals.

\emph{Energy selection strategies:} Twelve variants explored the influence of energy level inclusion by excluding specific energy values or using selected energy subsets such as 50–80–110~kVp or 60–90–120~kVp, to identify which energy levels are most critical for accurate model adaptation.

The spatial sampling patterns were selected empirically to ensure adequate coverage of the asymmetric beam intensity distribution observed in the Siemens system (10° anode tilt). These distances were determined based on preliminary profile measurements and the beam coverage on the detector. For systems with smaller anode angles where intensity asymmetry is less pronounced, the spatial sampling intervals may be re-optimized accordingly to maintain sufficient profile resolution.

\subsubsection{Effectiveness of fine-tuning versus training without prior knowledge}

To evaluate the effectiveness of fine-tuning compared to training a model without prior knowledge, we performed a comparative analysis between two models, the fine-tuned model \( M_2' \), which is adapted from the base model \( M_1 \) trained on measurements from the Siemens machine, and a model \( M_2 \) trained directly on Philips machine data. Both models yield predictions \( P_2' \) and \( P_2 \) respectively, which  were assessed against the corresponding experimental measurements obtained from the Philips system. The primary motivation for adopting a fine-tuning approach, rather than retraining a new model, lies in its efficiency. Fine-tuning allows us to leverage the knowledge already embedded in a pre-trained model, significantly reducing both computational cost and the number of required measurements, while still achieving prediction errors within the clinically acceptable range.

\subsection{Implementation into Monte Carlo softwares}

The proposed approach was implemented in OpenGATE 10\parencite{sarrut2022} and GGEMS\parencite{Bert2013}, the function \verb|source.direction.type='iso'|  was used to generate an isotropic, uniform source with no variation in the weights. To integrate our model into the MC toolkits OpenGATE and GGEMS, an existing function which was configured with \verb|source.direction.type='histogram'| was used to assign the predicted weights. This configuration is similar to 'iso', but it allows the emission of primary particles with directional distributions weighted by custom-defined histograms for $\theta$ and $\phi$ angles. The weights are specified using \verb|source.direction.histogram_phi_weights| at each energy level and for each angle defined via \verb|source.direction.histogram_phi_angles|.

The angle \(\phi\) was computed using the following geometric relationship:

\begin{equation}
    \phi = \arctan\left(\frac{D_{\mathrm{Lat}}}{SDD}\right)
\end{equation}

where \(D_{\mathrm{Lat}}\) denotes the lateral displacement of the dose measurement point from the beam central axis (isocenter), and \(SDD\) corresponds to the source-to-detector distance. In this context, the measurement plane refers to the plane orthogonal to the central X-ray beam axis where dose measurements are acquired, typically coinciding with the detector surface.

In addition, to transition between points along the anode-cathode axis, linear interpolation was applied to the weights. These weights were adjusted for the tube potentials used, thereby accounting for the energy-dependent nature of the anode heel effect. For energies not explicitly defined in these increments, the predicted weights from the trained model were employed.

\subsubsection{Evaluation on the fluence calculation}

A validation test was conducted to assess how the anode heel effect influences radiation distribution over a detector surface. The detector was placed 70 cm from the X-ray source to evaluate dose distribution. The resulting dose maps, generated with and without accounting for the anode heel effect, exhibited significant differences in intensity patterns. This further validated the accuracy of the AI model in capturing the phenomenon.

\subsubsection{Evaluation on the image quality} 
An anthropomorphic phantom PH-2E, a modified version of PBU-60 \parencite{kim2013study} with pathologies, was used from PLaTIMeD to simulate a patient scenario in both the experimental setup and the MC simulation, as shown in figure \ref{figure1_1}. The phantom represents a human with a height of 165 cm and a weight of 60 kg. An anterior-posterior (AP) X-ray of the upper chest was performed at 75 kVp using the Cios Alpha X-ray machine. The same parameters were applied in the MC simulation, first using an isotropic uniform X-ray beam distribution and then incorporating the anode heel effect into the beam. Finally, the percentage relative errors were calculated to evaluate the impact of the anode heel effect in producing a real-world modulated X-ray beam.

\section{Results}

\subsection{Characterization of the Anode Heel Effect from Experimental Measurements}
\label{Characterization of fthe anode heel effect}

\begin{figure}[]
    \centering
    \begin{subfigure}{0.495\linewidth}
        \centering
        \includegraphics[width=\linewidth]{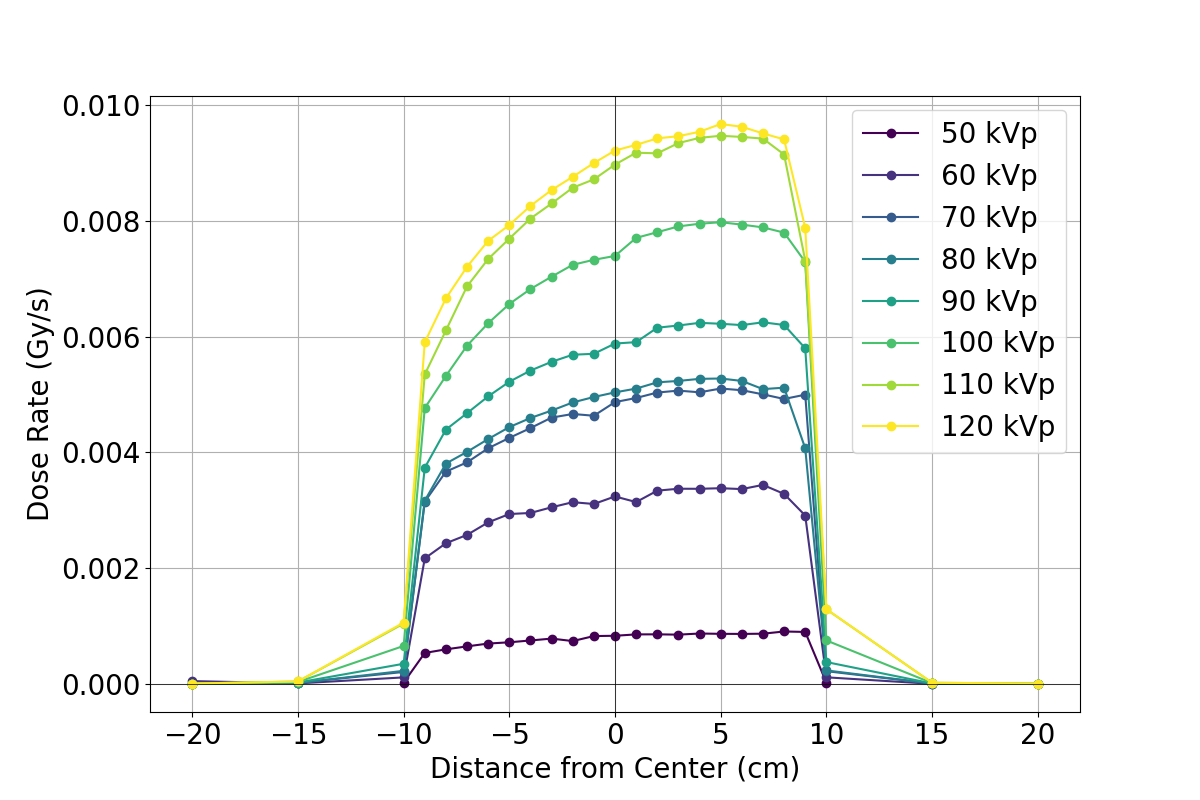}
        \caption{}
        \label{figure_2}
    \end{subfigure}
    \hfill
    \begin{subfigure}{0.495\linewidth}
        \centering
        \includegraphics[width=\linewidth]{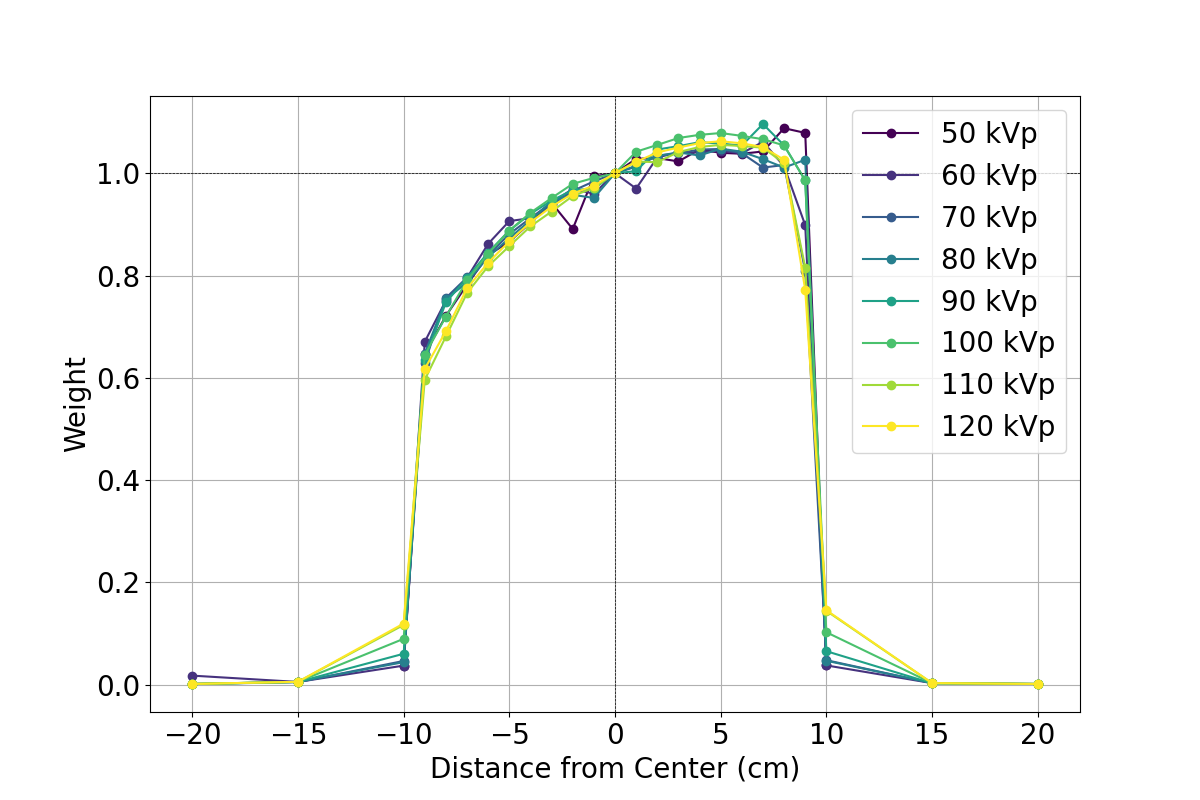}
        \caption{}
        \label{figure_3}
    \end{subfigure}
    \caption{Comparison of experimental measurements and their corresponding weights affected by the anode heel effect along the anode–cathode axis. (a) Dose value profiles as a function of distance from the X-ray beam center for various tube potentials. (b) Beam weight distributions illustrating the energy-dependent asymmetry caused by the anode heel effect.}
    \label{figure_2_3}
\end{figure}

Experimental measurements performed on the Siemens Cios Alpha X-ray system across tube potentials from 50 to 120~kVp revealed a pronounced and energy-dependent asymmetry in dose value distribution along the anode–cathode axis. As shown in Figure~\ref{figure_2}, dose values increased by approximately 9.6\% on the cathode side and decreased by up to 12.5\% on the anode side, particularly beyond 5~cm lateral displacement. These findings are consistent with the expected behavior of the anode heel effect and confirm its clinical significance.

The corresponding relative intensity weights, normalized to the central beam intensity, are shown in Figure~\ref{figure_3}. These weights consistently reflected an asymmetrical beam profile, forming the basis for the subsequent training and validation of the virtual source model.

While the anode heel effect itself is primarily determined by the anode angle and material, minor apparent variations with tube potential arise due to the changing energy spectrum and corresponding attenuation within the tungsten target. In this work, the variation of beam energy was intended to characterize a physical dependence, and to evaluate the robustness of the proposed model under different beam spectra used in clinical conditions.

\subsection{Machine Learning Model Performance for Heel Effect Prediction}

Table~\ref{table_1} summarizes the performance of the six machine learning models evaluated for predicting the spatial intensity distribution as a function of tube potential and dosimeter position. The GBR outperformed all other models, achieving a test MSE of 0.0014 and a corresponding $R^2$ score of 0.963. Decision Tree and Random Forest models also exhibited strong performance with $R^2$ values of 0.943 and 0.928, respectively.

\begin{table}[]
    \centering
    \caption{Performance comparison of different machine learning models for predicting the anode heel effect. Models are ordered by accuracy, highlighting the superiority of the Gradient Boosting Regressor (GBR) in capturing dose distribution asymmetries.}

    \begin{tabular}{>{\raggedright\arraybackslash}m{4cm} >{\centering\arraybackslash}m{4cm} >{\centering\arraybackslash}m{3cm} >{\centering\arraybackslash}m{3cm}}
        \toprule
        \textbf{Model} & \textbf{Mean Cross-Validation MSE} & \textbf{Test MSE} & \textbf{Test $R^2$} \\
        \midrule
        LinearRegression & 0.0333 & 0.0432 & 0.7909 \\
        MLP & 0.0118 & 0.0200 & 0.9030 \\
        SVR & 0.0102 & 0.0174 & 0.9159 \\
        RandomForestRegressor & 0.0039 & 0.0024 & 0.9284 \\
        DecisionTreeRegressor & 0.0042 & 0.0015 & 0.9429 \\
        \textbf{GradientBoostingRegressor} & \textbf{0.0019} & \textbf{0.0014} & \textbf{0.9631} \\
        \bottomrule
    \end{tabular}
    \label{table_1}
\end{table}

In contrast, classical Linear Regression yielded a test MSE of 0.0432 and a lower $R^2$ of 0.79, demonstrating its limited capacity to capture the nonlinear intensity gradients. Support Vector Regression and Multilayer Perceptron achieved moderate performance, with $R^2$ values near 0.91. These results highlight the superior generalization and fitting capabilities of ensemble-based models, particularly GBR, for heel effect modeling.

\subsection{Validation of GBR-Predicted Weights Against Experimental Data}

\begin{figure}[]
    \centering
    \makebox[\textwidth][c]{ 
        \includegraphics[width=0.95\textwidth]{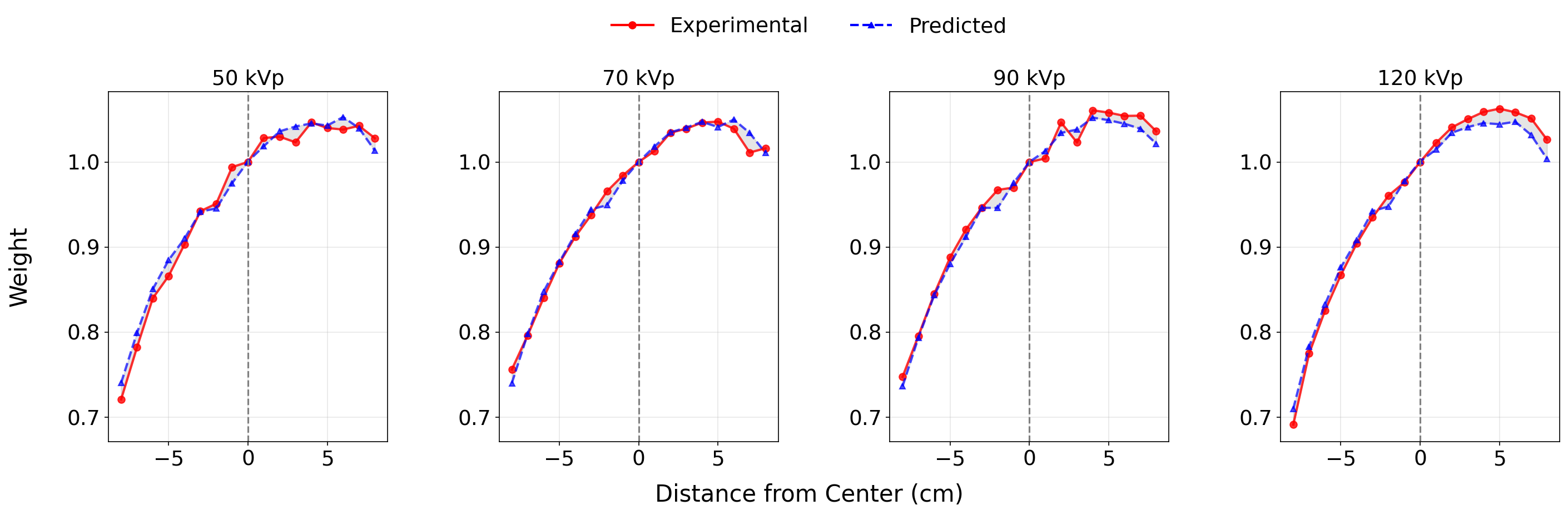}
    }
    \caption{Comparison between experimentally measured and AI-predicted weights from Model M1 (P1) across different tube potentials. The negative side corresponds to the anode, and the positive side to the cathode.}

    \label{figure_4}
\end{figure}

Figure~\ref{figure_4} presents a comparison between experimentally measured weights and those predicted by the GBR model for various tube potentials. Across all energy levels, the model predictions closely matched the experimental values, with relative percentage errors remaining below 5\%. The GBR model effectively captured the nonlinear asymmetry introduced by the anode heel effect and its dependence on beam energy, reinforcing its applicability for high-fidelity beam modeling.

\subsection{Evaluation of Fine-Tuning Strategies for Cross-Machine Adaptation}

To assess the feasibility of adapting the trained model to different X-ray systems, multiple fine-tuning strategies were evaluated using experimental data from the Philips Zenition 70 system. As summarized in table~\ref{table_2}, the most efficient strategy involved sampling six spatial positions per energy level (at $-8$, $-5$, $-2$, $+2$, $+5$, and $+8$~cm) across eight tube potentials. This protocol required only 48 measurements in total, which presents a 65\% reduction in data acquisition effort compared to full-field sampling, while still achieving a maximum prediction error of $1.37\% \pm 0.16\%$ and an $R^2$ score of 0.958.

\begin{table}
\caption{Comparison of fine-tuning strategies, ranked by $R^2$ score. The table presents the number of required measurement points and the corresponding mean percentage error (Error \%), demonstrating the trade-off between calibration effort and model accuracy.}
\centering

\begin{tabular}{l c c c}
\hline
\textbf{Strategy} & \textbf{Points} & \textbf{Error (\%)} & \textbf{$R^2$} \\
\hline
even distances	          & 72 &  1.36	& 0.959 \\

\textbf{6 points per energy}   & \textbf{48}  & \textbf{1.37} & \textbf{0.958} \\
8 points per energy   & 64  & 1.44 & 0.954 \\
odd distances          & 64  & 1.51 & 0.948 \\
quartiles               & 32  & 1.71 & 0.912 \\
4 points per energy   & 32  & 1.90 & 0.905 \\
center only            & 40  & 2.72 & 0.876 \\
edges only             & 48  & 2.08 & 0.863 \\
center extremes        & 24  & 3.56 & 0.702 \\
extremes                & 16  & 6.23 & 0.348 \\
\hline
\end{tabular}
\label{table_2}
\end{table}

Although slightly lower in accuracy than the denser even-distance strategy (72 points), this approach significantly reduced the clinical burden of model adaptation without compromising prediction quality. These results support the six-point-per-energy protocol as a practical and scalable strategy for efficient cross-machine adaptation.

\subsection{Comparison Between Fine-Tuned and Fully Trained Models}

\begin{figure}[]
    \centering
    \makebox[\textwidth][c]{ 
        \includegraphics[width=0.95\textwidth]{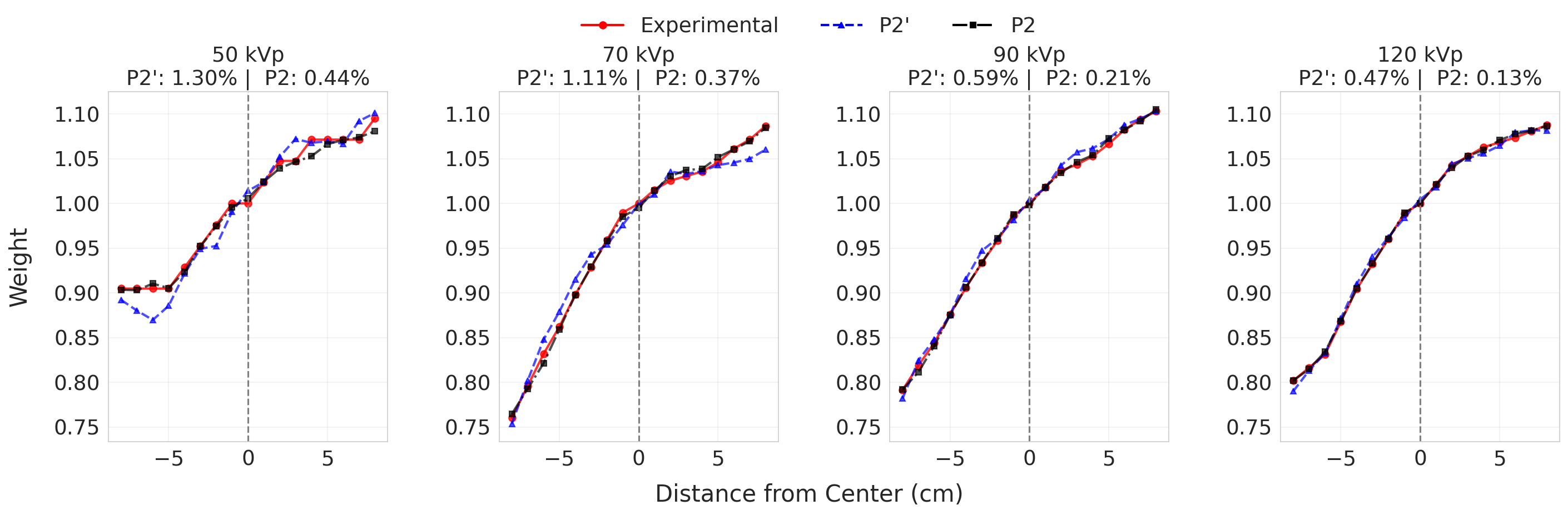}
    }
    \caption{Comparison of predicted weights from fine-tuned Model M2' (P2') and Model M2 (P2) against experimental weights, including percentage relative errors less than 1.5\%.}

    \label{figure_5}
\end{figure}

To further validate the proposed transfer learning methodology, we compared the predictions of a fine-tuned model (\(P_2'\)) against a fully trained model on Philips data (\(P_2\)). As shown in figure~\ref{figure_5}, both models achieved high accuracy, with \(P_2'\) deviating by a maximum of 1.38\% and \(P_2\) by only 0.57\% relative to the ground truth.

Despite the higher precision of the fully trained model, the fine-tuned approach required significantly fewer measurements and training resources. Specifically, while the full training used all 152 measurements, the fine-tuning strategy achieved comparable performance using only 48 measurements representing around 65\% reduction in required data. These results emphasize the potential of transfer learning in enabling rapid model deployment across varied clinical systems with minimal accuracy loss.

\subsection{Simulation-Based Evaluation: Fluence Maps and Image Quality}

\begin{figure}[]
    \centering
    \makebox[\textwidth][c]{ 
        \includegraphics[width=0.6\textwidth]{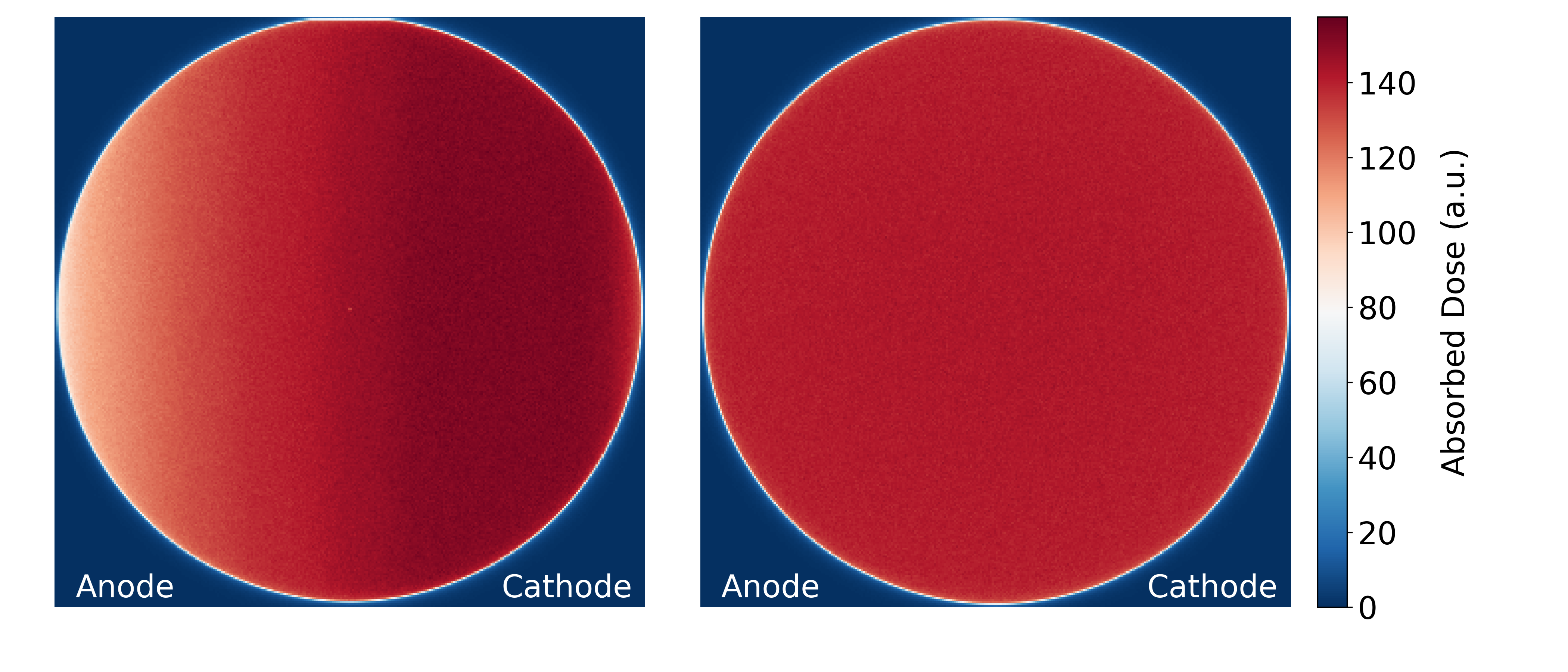}
    }
    \caption{Dose map validation comparing the AI-modeled anode heel effect (left) with the standard uniform emission model (right). The asymmetric dose distribution in the left panel highlights the impact of the anode heel effect, whereas the right panel illustrates a symmetric distribution resulting from uniform beam emission.}

    \label{figure_6}
\end{figure}

The impact of the anode heel effect on radiation distribution was further evaluated through Monte Carlo simulations which validate the implementation of the approach in OpenGate 10 and GGEMS. Figure~\ref{figure_6} compares dose maps generated with and without the anode heel effect using the implemented GBR-based virtual source. The inclusion of heel-effect-based weights produced visibly asymmetric fluence distributions, especially in peripheral regions, closely resembling measured profiles.

\begin{figure}[]
    \centering
    \includegraphics[width=0.9\linewidth]{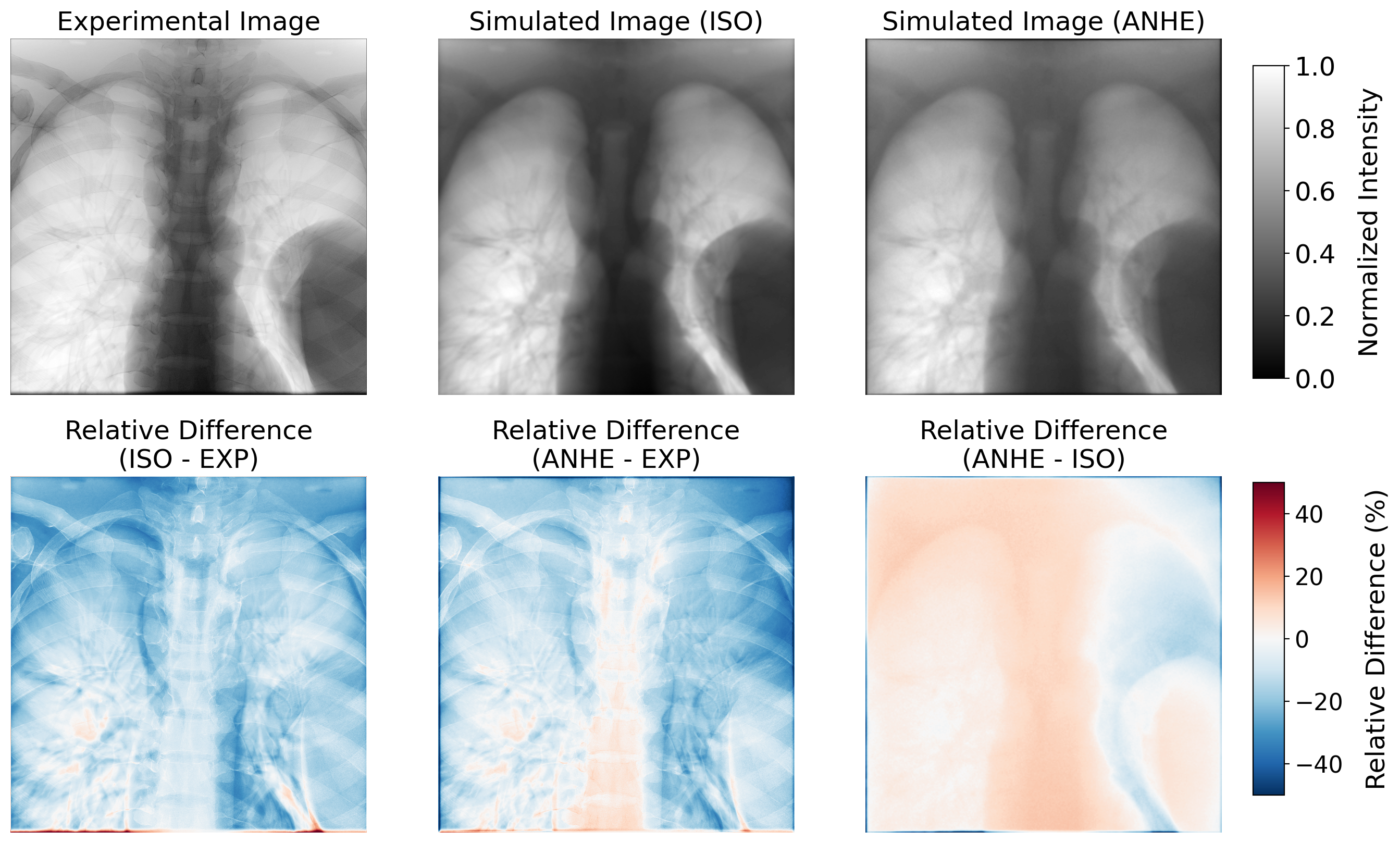}
    \caption{Top row shows the X-ray images of the same phantom: experimental (EXP), simulated with isotropic source (ISO), and simulated incorporating the anode heel effect (ANHE), all shown in grayscale. Bottom row shows the corresponding relative percentage difference maps: EXP vs. ISO, EXP vs. ANHE, and ISO vs. ANHE, highlighting areas of agreement and discrepancy between images.}

    \label{figure_7}
\end{figure}

Additionally, chest X-ray image simulations using an anthropomorphic phantom (PH-2E) were performed with the same parameters and $10^9$ beam photons. Figure~\ref{figure_7} shows that images generated with the AI-weighted source exhibited improved anatomical realism and edge fidelity compared to the isotropic model. Some differences were still observed when comparing our simulation source model image with the experimental one, which may be attributed to post-processing performed by the imaging system. The difference maps confirmed that incorporating the anode heel effect reduced structural underestimation, particularly in the lung regions. These results underscore the importance of integrating physically realistic beam models into simulation frameworks for accurate image synthesis and dosimetric studies.

\section{Discussion}

In this study, we introduced a hybrid approach combining experimental measurements, machine learning regression, and Monte Carlo simulations to accurately model the anode heel effect in simulated clinical X-ray systems. The proposed methodology aimed to address the limitations of conventional uniform source modeling by capturing the spatial and anode angle-dependent asymmetries inherent in real X-ray tubes. The evaluation across tube potentials primarily served to test the generalization capability of the model to energy-dependent spectra and to study the physical dependence of the heel effect on voltage.

The results demonstrated that the anode heel effect induces a consistent spatial dose asymmetry along the anode–cathode axis, with intensity reductions of up to 12.5\% on the anode side and enhancements nearing 10\% on the cathode side, as presented in Section \ref{Characterization of fthe anode heel effect}. These observations are in line with previous experimental findings reported by \parencite{shin2016anode} and validate the importance of modeling this phenomenon for accurate dose simulation. Importantly, the experimental weight distributions served as a reliable ground truth to train and evaluate predictive models.

Among the six regression algorithms tested, the Gradient Boosting Regressor achieved the best performance, accurately reproducing the heel effect behavior across all energy levels with a test MSE of 0.0014 and an \( R^2 \) score exceeding 0.96. These preliminary findings indicate that GBR seems to offer good generalization, requiring relatively few measurements for effective fine-tuning to different operating conditions. Nevertheless, further work is necessary to assess its generalizability across multiple fluoroscopy systems.

A key contribution of our work lies in the investigation of fine-tuning strategies to enable model adaptation across different X-ray systems. Notably, a reduced measurement protocol involving only six spatial points per energy level allowed for significant reductions in acquisition time (up to 65\%) while still achieving clinically acceptable prediction errors below 1.4\%. This demonstrates that transfer learning techniques can successfully adapt a model trained on one device (Siemens) to another (Philips), thereby promoting scalability and reducing resource burden in clinical commissioning scenarios. The comparison between the fine-tuned model \(P_2'\) and the directly trained model \(P_2\) confirmed that performance trade-offs remain minimal even with significantly fewer training data points, reinforcing the practicality of the proposed strategy. The model’s adaptability to variations in tube geometry or anode material is achieved through the fine-tuning strategy presented in Section \ref{Cross-machine fine-tuning strategies}. By retraining a small residual network on a limited number of measurements, the model effectively compensates for system-specific differences such as anode angle or target composition. This adaptability was validated between Siemens and Philips systems, suggesting strong generalization capabilities with minimal calibration data.

From a simulation standpoint, the integration of AI-predicted beam weights into OpenGATE 10 and GGEMS frameworks through histogram-based angular distributions allowed for physically informed particle direction sampling. This modification markedly improved the realism of fluence maps and image outputs. Validation using dose maps and clinical X-ray images illustrated the substantial differences between isotropic and anode-heel-aware models, particularly in peripheral beam regions and soft tissue contrast. These results highlight the clinical relevance of incorporating anode heel modeling into MC pipelines for both dose estimation and image quality assessment.

While first-principles Monte Carlo simulations of electron–anode interactions could, in theory, provide a complete description of the anode heel effect, such simulations are typically unfeasible for clinical X-ray systems. They require detailed, often proprietary, information on tube geometry such as focal spot dimensions, filament position, and housing filtration, and are computationally intensive, requiring billions of electron histories to reach stable statistics. For this reason, our model relies on experimentally measured dose distributions, which directly reflect the effective emission profile of the actual clinical system, serving as a practical and physically validated alternative.

In clinical settings, accurately modeling the anode heel effect enhances dosimetry precision, which is critical for procedures that require controlled radiation exposure. The AI-driven approach also supports research by streamlining the simulation of realistic X-ray beam characteristics. The current implementation applies specifically to reflective tungsten anodes typical of diagnostic X-ray tubes. Extension to transmission target configurations, such as those used in radiotherapy, would require adapting the angular emission formulation and energy spectrum to account for forward-directed photon emission. Future work will focus on testing the model on a third fluoroscopy system to further assess its generalizability across different X-ray machines. Additionally, we aim to extend this framework to computed tomography (CT) applications, where integrating scanner-specific beam models in place of traditional bowtie filters may offer new opportunities for dose optimization and improved image quality. These developments highlight the potential of physics-informed AI to bridge the gap between simulation fidelity and clinical applicability in the next generation imaging systems.

\section{Conclusion}

Our work presents a novel and efficient method for modeling the anode heel effect using a data-driven approach integrated into MC simulations. The combination of accurate GBR-based regression, reduced measurement strategies, and MC-compatible implementation provides a scalable and adaptable framework suitable for both research and clinical applications. By improving the fidelity of X-ray beam modeling, this methodology contributes to more accurate dosimetry, enhanced image realism, and greater efficiency in simulation-based studies within medical imaging and radiation physics.

\section*{Acknowledgements}
This work was partially funded by the French National Research Agency through the MoCaMed project (ANR-20-CE45-0025) and the Brittany Region. Our work was also supported by the pre-clinical experimental facility PLaTIMeD.

\section*{Usage Note}
The modeled X-ray source described in this article is freely available for use in research and simulation studies. It is compatible with tools such as OpenGATE 10 and can be accessed through \href{https://github.com/OpenGATE/opengate/blob/master/opengate/tests/src/test075_siemens_cios_alpha.py}{this implementation} for the Siemens Cios Alpha X-ray system. We kindly request that users cite this article as a reference when utilizing the model in their work. Proper citation ensures acknowledgment of the development efforts and supports the dissemination of this resource within the scientific community.


\printbibliography

\end{document}